

\input harvmac.tex
\noblackbox

\def\pmb#1{\setbox0=\hbox{#1}%
  \kern-.025em\copy0\kern-\wd0
  \kern.05em\copy0\kern-\wd0
  \kern-.025em\raise.0433em\box0 }

\def\hs{{\bf H}}
\def\rv{{\bf r}}
\def\rpep{{\bf r}_{\perp}}
\def\rvi{{\bf r}_i}
\def\bftau{\pmb{$\tau$}}
\def\grad{\pmb{$\nabla$}}
\def\gradperp{\pmb{$\nabla_{\perp}$}}
\def\zhat{{\bf {\hat{z}}}}
\def\lf{\lambda_f}
\def\lnf{\lambda_{nf}}
\def\lac{\lambda_{ac}}
\def\lplus{\lambda_1}
\def\lmin{\lambda_2}
\def\ll{\lambda}

\def\lab{\lambda_{ab}}
\def\lc{\lambda_{c}}
\def\xiab{\xi_{ab}}

\def\visco{\tilde{\eta}_l}
\def\bflam{\pmb{$\Lambda$}}
\def\qw{({\bf q},\omega )}
\def\bfq{{\bf q}}

\Title{}{\vbox{\centerline{AC Response of the Flux-Line Liquid}
     \vskip2pt\centerline{in High-$T_c$ Superconductors}}}

{\baselineskip= 16pt plus 2pt minus 1pt\centerline{LEE-WEN CHEN and
M. CRISTINA MARCHETTI}
\footnote{}
\smallskip\centerline{Physics Department}
\centerline{Syracuse University}
\centerline{Syracuse, NY 13244}
\medskip

\vskip .1in
We use a hydrodynamics theory to
discuss the response of a viscous flux-line liquid to an ac perturbation
applied at the surface of the sample. The theory incorporates
viscoelastic effects and describes the crossover between liquid-like
and solid-like response of the vortex array as the frequency of the
perturbation increases. A large viscosity from flux-line interactions
and entanglement leads to viscous screening of surface fields. As a
result, two frequency-dependent length scales are needed to describe
the penetration of an ac field. For large viscosities the
imaginary part of the ac permeability can exihibit, in addition to
the well-know peak associated with flux diffusion across the sample, a
new low-frequency peak corresponding to the transition from solid-like
to liquid-like behavior.

\Date{3/94}

\newsec{Introduction}

The study of the response of high-temperature (HTC) superconductors
in the mixed state to alternating magnetic fields and transport
currents provides direct information on the dynamics of magnetic flux
lines in these materials
\nref\acearly{J.R. Clem, H.R. Kerchner and S.T. Sekula, Phys. Rev.
B14, 1893(1976).}
\nref\koch{H.K. Olsson, R.H. Koch, W. Eidelloth, and
R.P. Robertazzi, Phys. Rev. Lett. {\bf 66}, 2661 (1991).}
\nref\yeh{N.-C. Yeh, et al., Phys. Rev. B{\bf 45}, 5654(1992);
D.S. Reed, N.-C. Yeh, W. Jiamng, U. Kriplani, and F. Holtzberg,
Phys. Rev. B47, 6146(1993); N.-C. Yeh, et al.,
Phys. Rev. B47, 8308(1993).}\nref\wu{D.-H. Wu
and S. Sridhar, Phys. Rev. Lett. {\bf 65}, 2074 (1990).}\refs{\acearly - \wu}.
Such ac measurements can be performed by
superimposing a small alternating magnetic field to the constant
applied field, $\hs$. Alternatively, one can measure the change in
resonance frequency of the superconducting crystal undergoing
mechanical vibrations \nref\gammel{P.L. Gammel,
L.F. Schneemeyer, J.V. Waszczak, and D.J. Bishop, Phys. Rev. Lett. {\bf 61},
1666 (1988).}\nref\farrell{D.E. Farrell, J.P. Rice, and D.M. Ginsberg,
Phys. Rev. Lett. {\bf 67}, 1165 (1991).}\refs{\gammel , \farrell}.
Due to the pinning of the flux lines to the crystal
lattice, the periodic tilting of the superconductor is equivalent to
the application of an ac field normal to $\hs$.

AC techniques are also
used to determine the irreversibility line, $T_{irr}(H)$, in disordered HTC
superconductors \nref\maloz{A.P. Malozemoff, T.K. Worthington,
Y. Yeshurun, and F. Holtzberg, Phys. Rev. B{\bf 38}, 7203 (1988).}
\refs{\maloz}.
In magnetization measurements this is defined as the locus in the $(H,T)$
phase diagram where the magnetization for field-cooled samples
differs from the result obtained in samples cooled initially at zero
field. This criterion is experimentally ambiguous since it depends
on the rate at which data are taken. For this reason ac techniques,
such as ac permeability, ac transport measurements, ac field
penetration, mechanical oscillator
\refs{\gammel , \farrell}, and others
are often preferable. When cooling through the
irreversibility line during an ac experiment the dissipation goes
through a maximum. Experimentally the location of this maximum is very
close to $T_{irr}(H)$ and is used as an alternative definition of
$T_{irr}$. The location of $T_{irr}(H)$ has great technological as
well as intellectual interset since it generally coincides with the
line below which the resistivity becomes unmeasurably small.

Much theoretical work has been done to investigate the high-frequency
electrodynamic response of a vortex solid pinned by weak impurity
disorder
\nref\koshelev{A.E. Koshelev and V.M. Vinokur, Physica C{\bf 173},
465 (1991).}\nref\coffey{M.W. Coffey and J. R. Clem, Phys. Rev. Lett.
{\bf 67}, 386 (1991); Phys. Rev. B {\bf 45}, 9872 (1992).}
\nref\brandt{E.H. Brandt, Phys. Rev. Lett. {\bf 67}, 2219 (1991).}
\nref\beek{C.J. van der Beek, V.B. Geshkenbein and V.M. Vinokur,
Phys. Rev. B {\bf 48}, 3393 (1993).}
\refs{\koshelev ,\beek}.
When the perturbing field is step-like in time, the
perturbation initially interacts with the flux line only at the
surface of the sample. The field, transport current or mechanical tilt
generate a surface current that flows in a layer of thickness of
the order of the static penetration length. This surface current
exerts a force on the vortices, deforming the vortex array. Flux-line
interactions cause the surface deformation to propagate in the
interior of the sample, while impurity pinning and friction tend to
impede its propagation.
Using standard descriptions of flux-line dynamics, one
finds that
for a periodic perturbation modulated at
frequency $\omega$, the penetration of the applied field
is governed by a single complex penetration length
$\lambda_{ac}(\omega )$ [1,8,9]. This length is closely related to the
skin depth in normal metals and determines directly the surface impedance
and the ac permeability of the material.
The peak in the ac dissipation occurs when
$\lambda_{ac}(\omega)$
is of the order of the sample size and is conventionally interpreted as
associated with the thermally activated depinning of the flux lines
\refs{\coffey ,\brandt}.
More recently the peak in the ac dissipation in disordered
superconductors has been interpreted as the signature of a continuous
transition from a high temperature flux-line liquid to a low
temperature vortex glass \nref\ffh{D.S. Fisher, M.P.A. Fisher, and
D. Huse, Phys. Rev. B{\bf 43}, 130 (1990).}\refs{\ffh}.
Good fits of ac
measurements in single crystals and in thin films of YBCO have been obtained
with the scaling relations of the phenomenlogical vortex glass model
\refs{\koch ,\yeh}.
In contrast the dissipation peak from mechanical oscillator data in clean
samples has been interpreted in the context of a melting of the flux-line
crystal to a
flux-line liquid \refs{\gammel ,\farrell}.

On the other hand, in a large part of the region of the $(H,T)$ phase
diagram probed by both dc and ac transport experiments the flux-line
array is in a liquid state. In this case it clearly no longer makes
sense to invoke the traditional description of flux-line dynamics
\nref\larkin {A.I. Larkin, Sov. Phys. JETP {\bf 31}, 784 (1970); A.I.
Larkin, and Yu N. Ovchinnikov, J. Low Temp. Phys. {\bf 34}, 409
(1979).}\refs{\larkin} in terms of the collective motion of
crystalline flux bundles. The question then arises of how to
incorporate collective effects in the description of flux-line
dynamics in the liquid phase.

A popular phenomenological approach for describing the dynamics of
flux-line liquids is the `` thermally activated flux-flow '' ( TAFF )
idea \nref\Kes{For a review, see P.H. Kes, J. Arts, J. van den Berg
and J. A. Mydosh, Supercond. Sci. Technol. {\bf 1}, 242
(1989).}\refs{\coffey , \beek, \Kes}, which essentially assumes an ideal gas of
disconnected flux elements moving in a tilted washboard potential.
This method generalizes the usual Bardeen-Stephen (BS) flux flow model
\nref\Kim{ See, for instance, Y.B. Kim, M.J. Stephen and W.F.
Vinen, in {\it Superconductivity}, vol. 2, R.D. Park, ed. (Marcel
Dekker, New York, 1969).}\refs{\Kim} to incorporate thermally-activated
depinning of the flux lines. At temperatures higher than a typical pinning
energy it reduces to the BS model.
The TAFF phenomenology has been
used to describe both dc \nref\copper{M. Inui, P.B. Littlewood,
and S.N. Coppersmith, Phys. Rev. Lett. {\bf 63}, 2421 (1989).}\refs{\copper}
and ac response \refs{\coffey}.
This picture takes, however, little account of intervortex
interactions in the flux liquid and is not very useful for describing a
system undergoing a phase transition, where collective effects can lead to
very large or diverging correlation lengths near the transition.

As discussed elsewhere, a useful framework for the description of the
long wavelength dynamical response of a flux-line liquid is
hydrodynamics \nref\mcmrev{M.C. Marchetti and D.R. Nelson, Physica C{\bf 174},
40 (1991).}\refs{\mcmrev}.
In this context intervortex interactions are naturally
incorporated in a viscosity that increases and goes to infinity at the
phase transition from a vortex liquid to a vortex crystal or glass. In
fact the behavior of the viscosity at the transition serves as a
signature of the order and nature of the transition itself.
The hydrodynamic model naturally incorporates the intrinsic nonlocality
of the electrodynamic response of a viscous flux liquid.
Marchetti and Nelson used hydrodynamics to describe
the response to dc perturbations \nref\mcma{M.C. Marchetti and D.R.
Nelson, Phys. Rev. B{\bf 41} (1990).}\refs{\mcma ,\mcmrev}.
They showed that a large viscosity,
as one can have as the phase transition is approached from the liquid
side, allows the effect of large scale spatial inhomogeneities, such
as macroscopic twin boundaries, to propagate large distances into the
flux liquid and slows down considerably the motion of the flux lines.
More recently Huse and Majumdar \nref\majum{D.A. Huse and S.N. Majumdar,
Phys. Rev. Lett. {\bf 71}, 2473 (1993).}\refs{\majum} used the hydrodynamic
model to describe multiterminal dc transport measurements in the vortex
liquid regime of the cuprate superconductors.
By calculating the dc response of a viscous flux liquid
for realistic geometries
including the effect of the sample boundaries, they demonstrated
that the
nonlocal resistivity arising from a finite flux liquid viscosity
can explain the voltage patterns seen in experiments
\nref\gammel{H. Safar {\it et al.} Phys. Rev. Lett. {\bf 72}, 1272
(1994).}
\refs{\gammel}.

In this paper we use the hydrodynamic model to describe the response of
a viscous flux-line liquid
to an ac perturbation applied at the surface of the sample.
A large viscosity can impede the penetration of a surface
field, since the viscous flux liquid cannot quickly adjust to the
external field change. This is reflected in a nonlocal relationship between
the vortex
liquid flow velocity, which determines the electric field from flux flow,
and the applied fields and currents. The nonlocality arises from viscous
forces, which are proportional to the second spatial derivatives of the
vortex flow velocity, and incorporates the force that remote fluid
elements exerts on each other via intervortex interactions and entanglement.
As a consequence of this nonlocality, the amplitude of the penetrating
ac field  is given by the linear superposition of two exponentially
decaying contributions with different frequency-dependent penetration
lengths, $\lplus$ and $\lmin$.
Recently Sonin et al.
\nref\sonin{E.B. Sonin, A.K. Tagantsev and K.B. Traito, Phys. Rev. B {\bf 46},
5830 (1992).}\refs{\sonin} showed that the penetration
an ac surface field that
generates a periodic tilt of the flux lines is also governed by two
characteristic length scales. In this case the new penetration length
is associated with an elastic tilt mode of the vortex array that
can impede surface penetration and was
neglected in previous theoretical studies.
In a viscous flux liquid for surface perturbations that compress or shear
the vortex array there is a viscous mode that incorporates
collective effects from intervortex interactions and entanglement
and is responsible for the additional surface screening.
This is to be contrasted to conventional TAFF or flux flow models,
where the resistivity is local and
the penetrating field has the form of a single exponential
controlled by the single length $\lambda_{ac}(\omega)$.
The largest of the two penetration lengths, $\lambda_1(\omega)$,
controls field penetration
in the bulk of the material. At low frequency it simply describes Meissner
screening. When the fluid viscosity is negligible
$\lambda_1(\omega)\approx\lambda_{ac}(\omega)$.
The additional penetration length, $\lambda_2(\omega)$,
is associated with viscous screening that prevents the magnetic field
from building up to its maximum value right at the surface.
The maximum of the penetrating field is then found at a distance
of the order of $|\lambda_2(\omega)|$ inside the sample.
Viscous screening is effective in a thin surface layer of
width $|\lambda_2(\omega)|$
($|\lambda_2(\omega)|<<|\lambda_1(\omega)|$ at all frequencies)
and leads to the nonlocality of the electric field from
flux motion. In a dense vortex array at low frequency
$|\lambda_2(\omega)|$ is of the order of the intervortex spacing.

In Section 2 we discuss the basic equations needed to describe
the electrodynamic response of a type-II superconductor in the
mixed state for the case when the flux array is in a liquid regime.
These are the usual coupled Maxwell and London equations
and a set of hydrodynamic equations for the viscous flux liquid.
In Section 3 we present the solution for an ac field penetrating
at the surface of a semiinfinite superconductor. The role of
viscosity and its effect on macroscopic properties, such as the
material's surface impedance are discussed. The solution for
a superconducting slab is given in Appendix A.

\newsec{Flux-Line Hydrodynamics}

We consider a uniaxial type-II superconductor in a
static field $\hs$ applied along the $c$ axis of the material,
which is chosen as the $z$ direction, i.e., $\hs=H\zhat$.
The static field
produces an array of flux lines that are on the average aligned with
the field direction. The mean spacing $a_0$ of the flux lines in the
$xy$ plane is determined by the average equilibrium induction ${\bf
B}_0=B_0\zhat$ that penetrates the sample, with
$a_0\simeq 1/\sqrt n_0$ and $n_0=B_0/\phi_0$ the areal density of
flux lines ($\phi_0=hc/2e$ is the flux quantum).  For $H >> H_{C1}$ and
neglecting demagnetizing effects, $B_0\simeq H$.
For high $\kappa$
superconductors, such as the HTC materials,
there is a large part of the $(H,T)$ phase diagram where
$\xiab<<a_0<<\lab$ (i.e., $H_{C1}<<B<<H_{C2}$), with $\lab$ and $\xiab$
the penetration and coherence lengths in the $ab$ plane, respectively.
Here we restrict
ourselves to this region. In addition, we are interested in properties
of the flux array on length scales larger than the mean intervortex
spacing where one can describe the system in terms of a few
hydrodynamic fields coarse-grained over several flux-line
spacings.

The electrodynamics of a type-II superconductor is described by
Maxwell's equations for the local fields
${\bf b}({\bf r},t)$ and ${\bf e}({\bf r},t)$,
\eqn\faraday{{\grad}\times {\bf e}+{1\over c}\partial_t {\bf b}=0,}
\eqn\amper{\grad\times {\bf b}={4\pi\over c}{\bf j}+{1\over c}
\partial_t {\bf e},}
with $\grad\cdot{\bf e}=4\pi\rho$ and $\grad\cdot{\bf b}=0$ \ref\foota{The
induction
${\bf b}$, rather than the field ${\bf H}$, appears in Eq. \amper\
because ${\bf j}$ here is the total current density including the
equilibrium response of the medium. For a discussion of this point,
see M. Tinkham, {\it Introduction to Superconductivity} (McGraw-Hill,
New York, 1975), p. 158-159.}.
These equations have to be supplemented with a constitutive relation for the
current density, ${\bf j}$. Following Coffey and Clem \refs{\coffey}, we use a
two fluid model and write ${\bf j}={\bf j}_n+{\bf j}_s$. The normal
current density ${\bf j}_n$ is specified by Ohm's law, ${\bf
j}_n=\sigma_{n}{\bf e}$, with $\sigma_{n}$ the electrical
conductivity of the normal metal.
The constitutive equation for the supercurrent ${\bf j}_s$ is
obtained by minimizing the Ginzburg-Landau free energy functional
for an anisotropic superconductor in the presence of vortices.
As customary, the anisotropy is incorporated in an effective mass
tensor. In the London approximation, where the magnitude of the
order parameter is assumed constant and only fluctuations in
the phase $\theta$ of the superconducting order parameter
are retained, we obtain
\eqn\london{ \bflam\cdot{\bf j}_s = -{c\over 4\pi}
({\bf A}-\phi_0\grad\theta),}
where $\bf A$ is the vector potential, with ${\bf b}=\grad\times{\bf A}$, and
$\bflam$ is a diagonal tensor with components $\Lambda_{xx}=\Lambda_{yy}
=\lab^2$ and $\Lambda_{zz}=\lc^2$.
Here $\lc$ is the penetration length along the $c$ direction, with
$\lc/\lab=\sqrt{m_c/m_{ab}}=p>>1$.
Taking the curl of Eq. \london\ and averaging over lengths
large compared to the intervortex spacing, one obtains the
London equation in the presence of vortices,
\eqn\londonb{\grad\times\bflam\cdot{\bf j_s}={c\over 4\pi}
  (-{\bf b}+\phi_0{\bf T}).}
Here ${\bf T}={\hat{\bf z}}n+\bftau$, with $n(\rv, t)$ and
$\bftau(\rv,t)$ the coarse-grained hydrodynamic density and tilt
fields of the
flux array, respectively. The microscopic density and tilt
fields are given by
\eqn\nmic{n_{mic}(\rv ,t)=\sum_{i=1}^N\delta(\rpep - \rvi (z,t)),}
and
\eqn\tiltmic{\bftau_{mic} (\rv ,t)=\sum_{i=1}^N{\partial\rvi (z,t)
\over\partial z}\delta(\rpep - \rvi (z,t)),}
where $\rvi(z)$ is the position of the $i$-th vortex line
in the $xy$ plane
as it wanders along the $z$ direction,
and $\rv = (\rpep ,z )$.
The tilt field describes the local deviation of an element of
flux-line liquid from the alignment with the $z$ direction.
The hydrodynamic fields are
obtained by coarse graining the
corresponding microscopic fields
over several vortex spacings.
The density and tilt fields are not independent, but are related by
the constraint that flux-lines cannot start nor stop inside the medium.
This requires
\eqn\contp{\partial_z n + \gradperp\cdot\bftau=0.}

Using Maxwell's equations one can eliminate the current from
Eq. \londonb\ in favor of the fields ${\bf b}$
and ${\bf e}$.
This can be done by multiplying
Eq. \amper\ with the tensor $\bflam$ from the left and then taking
the curl, with the result,
\eqn\genlond{\grad\times\bflam\cdot(\grad\times{\bf b}) + {\bf b}
=\phi_0{\bf T}
-{4\pi\sigma_{n}\over c^2}(\lab^2\partial_t {\bf b}
   -\lc^2\gradperp\times\zhat e_z)
- {1\over c}\partial_t({\lab^2\over c}\partial_t{\bf b}
 -\lc^2\gradperp\times\zhat e_z).}
The second term on the right hand side of
Eq. \genlond\ comes from the normal part of the current density.
The last term arises from the displacement current and is negligible
for all frequency $\omega << {c\over\lambda}\approx 10^{15}$Hz.
Since we are interested in experiments carried out at frequencies no
higher than microwave, we will neglect this term here,
even though it can be easily incorporated in the calculation.

To describe the response of the superconductor to a perturbing field
$\delta{\bf H}_a$ applied at the surface of the sample, one needs to
solve Eq. \genlond\ with appropriate boundary conditions on the field.
The
local field {\bf B} is, however, coupled to the vortex distribution
through the right hand side of Eq. \genlond\ and
the vortices in turn move in
response to changes in the field, as discussed for instance by Coffey
and Clem \refs{\coffey}.
Equation \genlond\ has to be supplemented with equations describing the
vortex dynamics. Here is where our work differs from that of other
authors. We
assume the flux-line array is in a liquid state and describe its
dynamics through the set of hydrodynamic equations for the density and
tilt field discussed elsewhere. The hydrodynamic
equations
consists of a continuity equation for the density,
\eqn\conth{\partial_t n + \gradperp\cdot {\bf j}_v=0,}
with ${\bf j}_v=n{\bf v}$,
and a continuity equation for the tangent field,
\eqn\stress{\partial_t \tau_\alpha +\partial_\beta j^{\tau}_{\alpha\beta}
=\partial_z j_{v\alpha},}
where $j^{\tau}_{\alpha\beta}$ are the components of the
tangent flux
tensor. This is a $2\times 2$ antisymmetric tensor and has therefore
only one independent component, $j_{xy}^{\tau}=\epsilon_{\alpha\beta}j^{\tau}
_{\alpha\beta}=-j_{yx}^{\tau}$.
At equilibrium $n=n_0$ and the tangent field vanishes,
$\bftau_0=0$, since the flux lines are on the average aligned with the
applied field.
In the following we will only
discuss the linear response of the flux liquid to
external perturbations of small amplitude
\nref\noteb{Experimentally a test for the applicability of linear
response theory is that the response is independent on the amplitude
of the perturbation.}\refs{\noteb}.
The hydrodynamic equations can then be linearized in
the deviations of the hydrodynamic
fields from their equilibrium values, $\delta n=n-n_0$ and
$\delta\bftau=\bftau$, with ${\bf j}_v\approx n_0{\bf v}$,
\eqn\contlin{\partial_t \delta n + n_0\gradperp\cdot {\bf v}=0,}
\eqn\stresslin{\partial_t \tau_\alpha +\partial_\beta j^{\tau}_{\alpha\beta}
=n_0\partial_z v_{\alpha}.}
As discussed in \refs{\mcmrev}, these equations have to be closed
with constitutive equations for the fluxes. Neglecting the Hall
current, and keeping only terms linear in the fluctuations from equilibrium,
the constitutive equations for ${\bf j}_v$ and $j^{\tau}_{xy}$ are given by,
\eqn\navier{-\gamma {\bf v} + \eta_s\nabla_{\perp}^2{\bf v}
   + \eta_b\gradperp (\gradperp\cdot {\bf v})
   +\eta_z\partial_z^2{\bf v}
  -{1\over c}{\bf B}_0\times{\bf j}=0,}
\eqn\tangentf{-\gamma_\tau j_{xy}^{\tau}
   +\eta_\tau \nabla_{\perp}^2j_{xy}^{\tau}
   +\eta_{\tau z} \partial_z^2j_{xy}^{\tau}
  +{n_0\over c}{\bf B}_0\cdot {\bf j} =0,}
where ${\bf j}=(c/4\pi)\grad\times\delta{\bf B}$, with $\delta {\bf B}=
{\bf B}-{\bf B}_0$.
The first terms in both Eqs. \navier\ and \tangentf\ describe the
drag on the flux lines arising from interaction with the crystal
lattice, with $\gamma$ and $\gamma_\tau$ friction coefficients per unit volume.
The contribution to $\gamma$ from interaction of the normal
core electrons with the underlying crystal lattice
can be approximated by the Bardeen-Stephen coefficient
$\gamma_{BS}(T,H)
=n_0\pi\hbar^2\sigma_n/2e^2\xi_{ab}^2$\refs{\Kim}.
Weak point pinning centers can be approximately incorporated in
a renormalized friction
by assuming $\gamma\simeq\gamma_{BS}e^{U_p/k_BT}$, where
$U_p$ is a typical pinning energy, as done in thermally activated
flux flow (TAFF) models \refs{\Kes}.
The friction coefficient $\gamma_\tau$ can be related to the relaxation
rate of an overdamped helicon and we estimate $\gamma_\tau\approx\gamma$.
The next three terms in Eq. \navier\ and the next two terms in Eq. \tangentf\
describe the viscous drag in the flux-line
liquid from intervortex interactions and entanglement. These effects are
incorporated in the viscosity coefficients: $\eta_s$, $\eta_b$ and $\eta_\tau$,
denoting the shear and bulk viscosity coefficients of an anisotropic
liquid, and
the ``tilt'' viscosities $\eta_z$ and $\eta_{\tau z}$
associated with velocity gradients
in the direction of the applied field.
Finally, the last two terms on the left hand side of both Eq. \navier\
and Eq. \tangentf\ represent reversible
forces on an element of flux-line liquid.
To linear order in the hydrodynamic densities these can be
written as
\eqn\revf{-{1\over c}{\bf B}_0\times{\bf j}=
  -{1\over n_0}\int d\rv'\big[\gradperp \hat{c}_L(\rv -\rv')\delta n(\rv')
  -\partial_z \hat{c}_{44}^L(\rv-\rv')\bftau(\rv')\big],}
\eqn\tangentxy{{n_0\over c}{\bf B}_0\cdot{\bf j}=
 \int d\rv' \hat{c}^L_{44}(\rv-\rv')
   \zhat\cdot[\gradperp'\times\bftau(\rv')].}
Here $\hat{c}_L(\rv)$ and $\hat{c}_{44}^L(\rv)$ are real space elastic
constants.
Their Fourier transforms,
$c_L({\bf q})$ and $c_{44}^L({\bf q})$, are the nonlocal compressional
and tilt moduli, respectively \mcmrev .
The liquid tilt modulus can differ from that of an elastic flux array
because of flux-line cutting.
We note that the nonlocality of the elastic constants is
automatically incorporated in our treatment.

When discussing the response of a viscous vortex liquid to an
ac perturbation one needs to modify Eqs. \conth\ -\tangentf\ to
include viscoelastic effects analogous to those present in entangled
polymer melts \nref\mcmmmm{M.C. Marchetti, J. Appl. Phys. {\bf 69},
5185 (1991).}\refs{\mcmmmm}. These effects will be important in the vortex
liquid in the region of temperatures just above the transition to a
solid phase because the viscosity gets very large in this region.
If velocity gradients are small, viscoelastic effects can be
incorporated following the theory of polymer dynamics
by modifying Eqs. \navier\ and \tangentf\ as
\eqn\navierv{-\gamma v_\alpha({\bf r},t) + \int_0^t dt'\int d{\bf r}'
 C_{\alpha i\beta j}({\bf r}-{\bf r}',t-t')
      \partial_i'\partial_j' v_\beta({\bf r}',t')
-{1\over c}[{\bf B}_0\times{\bf j}({\bf r},t)]_\alpha=0,}
and
\eqn\tangv{-\gamma_\tau j_{xy}^\tau({\bf r},t)
  +\int_0^t dt' \int d{\bf r}'
C_{ij}^\tau({\bf r}-{\bf r}',t-t')\partial_i'\partial_j'
j_{xy}^\tau ({\bf r}',t')
  +{n_0\over c}{\bf B}_0\cdot{\bf j}({\bf r},t)=0,}
where latin indices run over the three cartesian coordinates,
$i=x,y,z$, while Greek indices only run over the two coordinates
normal to the applied field, $\alpha=x,y$. The components of the tensors
$C_{\alpha i\beta j}({\bf r},t)$ and $C_{ij}^\tau({\bf r},t)$
are the Green-Kubo correlation
functions that determine the viscosity coefficients.
It is convenient to introduce the Fourier and Laplace transforms of
these correlation functions, defined as
\eqn\fourier{\tilde{C}_{\alpha i\beta j}\qw =
\int_0^\infty dt\int d{\bf r}e^{i {\bf q}\cdot{\bf r} }
\tilde{C}_{\alpha i\beta j}({\bf r},t)}
for $Im(\omega) < 0$.
There are five independent Green-Kubo correlation functions,
defined by
\eqn\kubov{\eqalign{\tilde{C}_{\alpha i\beta j}\qw=
& \tilde{\eta}_s\qw\delta_{\alpha\beta}
  [\delta_{ij}-\delta_{iz}\delta_{jz}]
  +\tilde{\eta}_z\qw \delta_{\alpha\beta}\delta_{iz}\delta_{jz} \cr
 & +{1\over 2} [\tilde{\eta}_l\qw
-\tilde{\eta}_s\qw ][\delta_{\alpha i}\delta_{\beta j}
   +\delta_{\alpha j}\delta_{\beta i}],}}

and
\eqn\kubot{\tilde{C}^\tau_{ij}\qw =\tilde{\eta}_\tau\qw\delta_{\alpha\beta}
  [\delta_{ij}-\delta_{iz}\delta_{jz}]
  +\tilde{\eta}_{\tau z}\qw \delta_{\alpha\beta}\delta_{iz}\delta_{jz},}
where $\tilde{\eta}\qw$ for $\nu=s,z,l,\tau,\tau z$ are the frequency-
and wavevector-dependent viscosity coefficients.
These determine the static viscosity coefficients according to,
\eqn\green{\eta_\nu=\tilde{\eta}({\bf q}=0,\omega=0),}
for $\nu=s,z,l,\tau,\tau z$, where $\eta_l=\eta_s+\eta_b$ is
the longitudinal viscosity.

In entangled polymers the relaxation of shear and compressional
stresses described by the Green-Kubo correlation functions is
nonexponential \nref\doi{See, for
instance, M. Doi and S.F. Edwards, {\it The Theory of Polymer Dynamics}
(Oxford University, New York, 1986).}\refs{\doi}. This reflects
the existence of a wide distribution of relaxation times in the
systems. The same feature is expected to occur in strongly
interacting flux-line liquids. Here for simplicity we model
the time decay of the Green-Kubo correlation functions as exponential
according to the phenomenological Maxwell model of viscoelasticity
\nref\boon{See, for instance, J.P. Boon and S. Yip, {\it Molecular
Hydrodynamics} (McGraw-Hill, New York, 1980).}\refs{\boon}.
The wave vector and frequency-dependent viscosities are then given by
\eqn\viscom{\tilde{\eta}_\nu\qw={\eta_\nu\over 1+i\omega\tau_\nu({\bf q})}.}
The characteristic relaxation times $\tau_\nu$ are chosen so that
the viscoelastic model incorporates the essential feature that the
fluid behaves like a viscous liquid on long times scales and
as an elastic solid on short time scales. In other words for
$\omega>>1/\tau_\nu$ for all $\nu$, the hydrodynamic equations
should reduce to the equations of continuum elasticity for a
flux-line array. This requires
\eqn\timesh{\tau_s=\eta_s/G,}
\eqn\timecomp{\tau_l=\eta_l/(c_{11}(\bfq)+G-c_L(\bfq)),}
\eqn\timetilt{\tau_z=\eta_z/(c_{44}(\bfq)-c_{44}^L(\bfq)),}
where $c_{11}$, $G$ and $c_{44}$ are the compressional, shear and tilt
elastic moduli of a flux-line elastic medium.
For an Abrikosov flux-line lattice $G\approx c_{66}$ and the
wavevector dependence of the shear modulus $c_{66}$ is always
negligible. The compressional modulus is given by
$c_{11}(\bfq)=c_L(\bfq)+c_{66} $ and $c_{11}(\bfq)\approx c_L(\bfq)$
since $c_{66}<<c_L(\bfq)$ at all but the largest wavevector $q\sim
k_{BZ}$ with $k_{BZ}=\sqrt{4\pi/a_0}$ the Brillouin Zone boundary,
where $c_L(k_{BZ})\sim c_{66}$. As a consequence we can approximate
$c_{11}$ by $c_L$ in eq. (2.25) with the result $\tau_l\sim\eta_l/G$.
The wavevector dependence of both the shear and compressional
relaxation time is therefore negligible.

\newsec{Surface Impedance and AC Permeability}

In this section we discuss the response of a flux-line liquid to a weak
ac field $\delta{\bf H}_a=\zhat\delta H_a e^{i\omega t}$ applied
parallel to the static field $\hs$, with $\delta H_a<<H$.
For clarity we first consider
the case of a semiinfinite superconductor occupying the half space
$y\geq 0$. The perturbing field is applied at the surface, $y=0$,
and generates a surface current in the $+x$ direction.
The surface current in turn exerts a Lorentz force normal to the sample
boundary that yields a compression
of the flux array.
To evaluate the change in induction in the superconductor as a result
of this perturbation we need to solve the coupled set of Maxwell
and London equations and the hydrodynamic equations of the flux liquid.
Since the lines remain on the average aligned with the $z$ direction,
the change in induction will be of the form $\delta{\bf B}=\zhat\delta
B_z(y)e^{i\omega t}$ and $\bftau=0$. The linearized hydrodynamic equations
reduce to
\eqn\contdens{i\omega\delta n(y)+n_0{d v_y\over dy}=0,}
\eqn\veloc{-\gamma v_y(y)+\tilde{\eta}_l(\omega){d^2v_y\over dy^2}
        -{B_0\over c}j_x(y)=0,}
where $j_x(y)=(c/4\pi)(d\delta B_z/dy)$ is the total current density,
including the response of the medium. The field $\delta B$ includes
both the Meissner response to the applied field and the change in
induction generated by the motion of the vortices.
It is determined by the London equation \genlond . For the simple geometry
considered here this becomes,
\eqn\londonc{-{d^2\delta B_z\over dy^2}
   +\Big({1\over\lambda^2}+{1\over\lambda^2_{nf}}\Big)\delta B_z=
           {\phi_0\over\lambda^2}\delta n(y),}
where $\lnf=\sqrt{c^2/4\pi i\omega\sigma_n}$ is the normal fluid skin depth
and $\lambda$ is the London penetration length in the $ab$ plane.
For simplicity we drop in this section the subscript on $\lambda_{ab}$,
since this is the only penetration length that is relevant for the
geometry considered. Equation \londonc\ has to be solved
with the boundary condition $\delta B_z(y=0)=\delta H_a$.

When $\visco=0$, corresponding to a simple flux flow model in the absence of
pinning, one can eliminate $\delta n$ between Eqs. \londonc ,
\contdens\ and \veloc\ to obtain a second order differential equation
for $\delta B_z$, which has solution
\eqn\bfzero{\delta B_z(y)=\delta H_ae^{-y/\ll_{ac}(\omega)},}
where
\eqn\penone{\ll_{ac}(\omega)=\Big[{\ll^2+\lf^2(\omega)\over
  1+\ll^2/\lnf^2(\omega)}\Big]^{1/2},}
with $\lambda_f(\omega)=\sqrt{c_L(0)/i\omega\gamma}$ the ac penetration
length of a flux liquid \nref\gesh{V.B. Geshkenbein, V.M. Vinokur
and R. Fehrenbacher, PRB. {\bf 43}, 3748 (1991).}\refs{\gesh}.
As in a normal metal,
the response of the material to an ac field is entirely characterized by
a single length $\ll_{ac}(\omega)$, that plays the role of the skin
penetration depth in metals.
The inclusion of the normal current guarantees that $\ll_{ac}(\omega)$
reduces to the normal metal skin depth at $H_{c2}$ where $\ll$ diverges
\refs{\coffey}.
The frequency dependence of $\lambda_{ac}$ is controlled by
two frequency scales. The first,
$\omega_f=c_L(0)/\gamma\ll^2$ is associated with flux
diffusion: $\omega_f^{-1}$
represents the time it takes for the field to diffuse
a length $\ll$ into the sample.
The second frequency scale is the frequency governing ac field screening
by the normal fluid component, $\omega_{nf}=c^2/(4\pi\sigma_n\ll^2)$.
If the friction coefficient $\gamma$ is approximated by its Bardeen-Stephen
value, we find $\omega_f/\omega_{nf}=B/H_{c2}<1$.
The ratio of these two frequencies is even smaller if $\gamma$ is
renormalized to incorporate the pinning by weak point defects as
in TAFF models, since in that case
$\omega_f/\omega_{nf}\approx (B/H_{c2})\exp(-U_p/k_BT)$.
The frequency dependence of $\ll_{ac}(\omega)$
is characterized by three regions, as shown in Fig. 1.
For $\omega<<\omega_f$,
$\ll_{ac}(\omega)\approx\lf(\omega)\sim\omega^{-1/2}$ and
the field penetration is controlled
by flux flow.
For $\omega_f<<\omega<<\omega_{nf}$,
$\ll_{ac}(\omega)\approx\ll$ and one has essentially static
Meissner screening since the flux lines cannot flow on experimental
time scales. Finally, for $\omega>>\omega_{nf}$,
$\ll_{ac}(\omega)\approx\lnf(\omega)\sim\omega^{-1/2}$
and the electrodynamic response
is controlled by the normal fluid. We remark that the frequency
dependence of $\ll_{ac}(\omega)$ in the region where penetration is
controlled by flux flow is the same as in a normal metal.
In the flux liquid model considered here,
weak point disorder decreases $\omega_f$ and
widens the region of Meissner screening.
As pointed out by Geshkenbeim et al \refs{\beek}., $\omega_f$
is very large (typically $\omega_f\sim 10^{14}Hz$) and
in the frequency range where the experiments are usually carried
out ($1-10^6 Hz$) one has $\lac\approx\lf$.

The quantity that is usually measured at these frequencies is the
surface impedance $Z_s(\omega)=R_s+iX_s$, defined as
\eqn\surfimp{Z_s(\omega)={E_x(0)\over \int_0^{\infty}j_x(y) dy}
={4\pi\over c} {E_x(0)\over \delta B_z(0)},}
where $E_x(0)$ is the electric field at the surface.
The surface resistance, $R_s$, and the surface reactance, $X_s$,
determine the loss and the resonance frequency, respectively,
when the superconductor forms part of a resonant circuit.
When $\visco=0$ the surface impedance is given by the familiar
expression,
\eqn\surfimpf{Z_s(\omega)={4\pi i\omega\over c^2}\ll_{ac}(\omega).}
It is related to the ac resistivity, $\rho_{ac}(\omega)=E_x(y)/j_x(y)$,
by $Z_s(\omega)=[4\pi i\omega\rho_{ac}(\omega)/c^2]^{1/2}$.
If $|\lf|>>\ll$, the ac resistivity is
independent of frequency and it is simply given by the Bardeen-Stephen
dc resistivity, $\rho_{ac}=\rho_f=4\pi c_L(0)/c^2\gamma$ \refs{\gesh}.

Finally, for a superconducting slab of thickness $2W$
occupying the region of space $|y|\leq W$, one
can use this simple flux-flow model that neglects
intervortex interaction to evaluate the ac permeability,
$\mu(\omega)$, defined as
\eqn\suscep{\mu(\omega)={1\over 2W\delta H_a}\int_{-W}^{W}
  dy \delta B_z(y),}
with the result,
\eqn\suscepff{\mu(\omega)={\lambda_{ac}\over W}\tanh\Big(
   {W\over \lambda_{ac}}\Big).}
The ac permeability depends on the sample thickness $W$ through
\eqn\lacw{{\lambda_{ac}\over W}=
\big[{\lambda^2/W^2+\omega_D/i\omega\over
1+i\omega/\omega_{nf}}\big]^{1/2}}
where $\omega_D=\omega_f\lambda^2/ W^2$ is the inverse of the time for
flux diffusion across the sample. The normal fluid contribution is
negligible provided $\omega<<\omega_{nf}$, where $\omega_{nf} >>
\omega_D$. In this range of frequencies a crossover in $\mu^{'}$ and a
peak in $\mu^{''}$ occur when $|\lambda_{ac}|\sim W$ or
$\omega\sim\omega_D$, i.e., the time modulation of the applied field
matches the time for flux diffusion across the sample thickness. If
$\omega << \omega_D$, $|\lambda_{ac}| >> W$ and the field penetrates
completely, yielding $\mu^{'}=1$. If $\omega >> \omega_D$ the flux
liquid cannot diffuse on the time scale over which the perturbation
changes and the applied field can only penetrate a surface layer of
width $\lambda$, yielding $\mu^{'}\approx \lambda/W \tanh(W/\lambda) <<
1$. The crossover between these two limits is marked by a maximum in
the absorption $\mu^{''}$. Finally, for $\omega\sim\omega_{nf}$ the
normal fluid contribution becomes important and leads to a second
crossover from $\mu^{'}\approx \lambda/W \tanh(W/\lambda)$ for $\omega_D
<< \omega << \omega_{nf}$ to $\mu^{'}\sim 0$ for $\omega >>
\omega_{nf}$, accompanied by a second peak in $\mu^{''}$.
When $\omega >> \omega_{nf}$ the normal quasiparticles cannot follow
the changes in the external field which is completely expelled from the
sample. If $W >> \lambda$, both $\mu^{'}$ and $\mu^{''}$ are very small
at $\omega\sim\omega_{nf}$, where the normal fluid contribution
becomes dominant. As a result, the second peak in $\mu^{''}$ is very
small, as show in Fig. 2a, where $W/\lambda =10$. On the other hand,
if $W\geq\lambda$ the second peak in $\mu^{''}$ at
$\omega\sim\omega_{nf}$ can exceed the peak from flux flow ( Fig.
2b). In general, the frequencies $\omega_{nf}$ is, however,
outside the range of frequency usually probed in experiments.

Recently Coffey and Clem incorporated flux creep and pinning
in the model described above by introducing a frequency-dependent
single-vortex mobility (the inverse of our friction coefficient)
\refs{\coffey}. They found that, while pinning and creep affect the
location of the peak in the absorption, no new length scale
appears in the electrodynamic response, which is qualitatively
similar to that obtained from simple flux-flow models.

In a viscous flux liquid intervortex interactions and entanglement
as described by
the viscous force in Eq. \veloc\ make the electrodynamic response
nonlocal. This nonlocality is governed by the new viscous length
\eqn\visl{\delta(\omega)=\sqrt{\tilde{\eta}(\omega)/\gamma}}
that controls the
relative importance of the first two terms of the left hand side
 of Eq. (3.2). To
clarify the meaning of this new length scale it is useful to
eliminate the flux density and flow velocity from Eq. (3.1)-(3.3) in
favor of the electrodynamic fields $\delta B_z(y)$ and
$E_x(y)={B_0\over c}v_x(y)+{4\pi\lambda^2i\omega\over c}j_s$, with the
result,
\eqn\exjx{E_x-\delta^2{d^2E_x\over
dy^2}=\rho_fj_x(y)+{4\pi\lambda^2i\omega\over
c}\big[j_s-\delta^2{d^2j_s\over dy^2}\big],}
where $\rho_f=B_0^2/c^2\gamma$ is the flux flow resistivity, and
\eqn\fourth{{\lambda^2\delta^2\over 1+\ll^2/\lnf^2}{d^4\delta B_z\over dy^4}
-(\lambda^2_{ac}+\delta^2){d^2\delta B_z\over dy^2}
+\delta B_z=0.}
The second term on the right hand side of \exjx\ arises from Meissner
screening of the fields over a surface layer of width $\lambda$.
To understand the physical meaning of the new length scale
$\delta(\omega)$ we define a contribution from flux motion
, $E_x^f$, to the electric field by $E_x=E_x^f+{4\pi\lambda^2i\omega\over
c}j_s$. Substituting this into \exjx\ , we obtain
\eqn\exf{E_x^f-\delta^2{d^2E_x^f\over dy^2}=\rho_fj_x(y).}
When the viscosity is very small the second term on the left hand
side of Eq. \exf\ is always negligible and the field from flux
motion is simply determined by Ohm's law for a normal metal of
resistivity given by the flux flow resistivity, $\rho_f$. The
resistive response of the medium is local and the ac resistivity
$\rho_{ac}$ is independent of the frequency, with $\rho_{ac}=\rho_f$.
When $\delta$ is sufficiently large the resistive response of the
medium is nonlocal :
the electric field at ${\bf r}$ is determined by the current at
spatially remote points ${\bf r}'$ as a result of the force that
remote fluid elements can exert on each other via interactions and
entanglement.

Using the Maxwell model of viscoelasticity given in Eq. (2.23), the
frequency dependent viscous length $\delta(\omega)$ is given by
$\delta(\omega)=\delta_0(1+i\omega\tau_l)^{-1/2}$ where
$\delta_0=\sqrt{\eta_l/\gamma}$ is the static viscous length discussed
earlier by Marchetti and Nelson \refs{\mcmrev}.
The static viscosity of a flux-line liquid has been estimated elsewhere
\nref\cates{M. Cates, PRB, {\bf 45}, 12415 (1992).}employing analogies
with the physics of entangled polymer melts\refs{\mcmrev ,\cates} . Assuming
$\gamma\approx\gamma_{BS}$, one finds
$\delta_0\approx a_0\exp(U_\times/2k_BT)$,
where $U_\times(H,T)$ is the typical
energy barrier for flux-line cutting, which is expected to vanish at
$H_{c2}$. If the barriers to flux cutting are sufficiently large,
at low temperatures the vortex array can get stuck in a polymer glass
regime of entangled lines, characterized by infinite viscosity on
experimental time scales.
A simple estimate of the crossing energy gives
$U_\times\approx 2(\sqrt 2-1)\sqrt{m_{ab}\over
m_c}a_0\epsilon_0\ln\kappa$,
with $\epsilon_0=(\phi_0/4\pi\lambda_{ab})^2$
\nref\ucross{ S. Obukhov and M. Rubinstein, Phys. Rev. Lett. {\bf 66},
2279 (1991); D.R. Nelson, in `` Phase Transitions and Relaxations in
Systems with Competing Energy Scales'', NATO-ASI series Vol. {\bf 415}
( Kluwer Academic, 1993 ).}\refs{\ucross}.
This can also be written in terms of the clean flux lattice melting
temperature $T_m=\alpha_L^2\epsilon_0 a_0\sqrt{m_{ab}\over m_c}$, with
$\alpha_L\approx 0.15\sim 0.3$ the Lindemann parameter \refs{\ucross},
as $U_\times/T_m=c_\times/\alpha_L^2$, with $c_\times=2(\sqrt 2
-1)\ln\kappa$.
This shows that the crossing barrier associated with entanglement can
become very large above $T_m$ and preclude
crystallization on experimental time scales.
Recent calculations of the energy $U_\times$ by Moore and Wilkin
\nref\moore{ M.A. Moore and N.K. Wilkin, preprint (1994). }\refs{\moore}
and by Carraro and Fisher
\nref\carraro{ C. Cararro and D.S. Fisher, preprint (1994). }\refs{\carraro}
confirm these simple estimates. Using $\alpha_L\sim 0.3 $ and
$\kappa\sim 200$, we find $\delta_0\sim a_0e^{50T_m/T}$. It is then
clear that the static viscous length $\delta_0$ can become very large
in the flux liquid regime and the resulting nonlocality of the dc
response can be probed experimentally.

Viscous effects introduce two new frequency scales in the ac response
of a flux array. The first is the frequency
$\omega_l=1/\tau_l=G/\eta_l$ describing the relaxation of shear
stresses and controlling
the frequency dependence of the viscous length
$\delta(\omega)$. The second frequency scale is defined by
$|\lambda_f(\omega)|\sim\delta_0$, corresponding to
$\omega\sim\omega_\eta =c_L(0)/\eta_l$.
Both $\omega_l$ and $\omega_\eta$ decrease with
increasing static viscosity and in flux arrays $\omega_l
<<\omega_\eta$ since $G << c_L(0)$, as discussed below. The frequency
$\omega_l$ only enters through the frequency dependence of
$\delta(\omega)$ and governs the crossover
from liquid-like to solid-like response of the flux array as a
function of the frequency $\omega$ of the external probe. To
understand the crossover it is useful to first neglect the fourth
order derivate of the field in \fourth\ . The single length scale
governing ac field penetration is then
$\lambda_{ac}^2+\delta^2=\lambda^2+\lambda_f^2+\delta^2.$ At low
frequency $(\omega << \omega_l)$, $\delta(\omega)\approx\delta_0$ and
$\lambda_{ac}^2\approx\lambda^2+\lambda_f^2+\delta_0^2$, with
$\lambda_f=\sqrt{c_L(0)/i\omega\gamma}$. The model then simply describes
penetration via flux diffusion (flux flow) of a vortex liquid of
static viscosity $\eta_l$. In this case the viscous force
provides an additional static damping of the penetration field. This
additional damping is, however, negligible if $\omega < \omega_\eta $.
If $G << c_L(0)$, then $\omega_l << \omega_\eta$ and the
flux flow contribution to the
penetration length always dominates the contribution from $\delta_0$
in this low frequency regime.
At high frequency ( $\omega >> \omega_l$ ),
$\delta(\omega)\sim\sqrt{G/i\omega\gamma}$ and
$\lambda_{ac}^2+\delta^2\sim\lambda^2+{c_L(0)\over
i\omega\gamma}+{G\over i\omega\gamma}=\lambda^2+{c_{11}(0)\over
i\omega\gamma}$. In this regime  field penetration is governed by flux
flow of a vortex lattice. The model therefore describes the
crossover from liquid-like to solid-like response at the frequency
$\omega_l$. This frequency decreases as the static
viscosity increases and $\omega_l/\omega_f=(G/
c_L)(\lambda^2/\delta_0^2)$. Assuming $G\approx c_{66}$,
where $c_{66}$ is the shear modulus of the Abrikosov flux lattice,
we find $G/c_L(0)\approx(1-b)^2/4\kappa^2b^2$, if $b=B/H_{c2}>0.25$,
and $c_{66}/c_L(0)\approx 1/4\kappa^2b^2$
if $0.3/\kappa^2<b<0.25$.
\nref\shearmd{ E.H. Brandt, Phys. Stat. Sol. (b), {\bf 77}, 551
(1976).}\refs{\shearmd}
Using material parameters of YBCO at
$T=85 K$, we estimate $G/c_L(0)\leq 10^{-3}$
\nref\malter{ An alternative estimate of the shear modulus $G$ of
a polymer glass  has been given in [17] ,
$G\sim k_BT/ a_0^2\xi_z$, where $\xi_z\approx a_0^2\tilde{\epsilon}_1/k_BT$,
with $\tilde{\epsilon}_1=\epsilon_0(m_{ab}/m_c)\ln\kappa$ the flux-line
tilt energy, is the typical distance between entanglements
when traced along one line. This gives
$G/c_L(0)\approx4\pi(k_BT)^2/\tilde{\epsilon}_1\phi_0^2\approx 5\times
10^{-17}\AA^{-2}
K^{-2}~T^2\lambda^2$, which is again very small at all
temperatures of interest here.}\refs{\shearmd,\malter}.
As a result,
at the frequency $\omega\sim\omega_l$ ( when the crossover
from liquid-like to solid-like behavior occurs )
the conventional flux  flow contribution to the penetration length is
so large that it always dominates the ac response of the flux array.
This result is not unexpected. It simply reflects the fact that the
only difference in response between a vortex solid and a vortex liquid
arises from the small difference in the compressional moduli.
A perturbation that generates a compression of the
flux array is therefore  not a good probe to
distinguish between liquid-like and solid-like response.
Viscous effects may be observable only provided $G/c_L\sim 1$, as
discussed below.

We now discuss in detail the field penetration for the semiinfinite geometry
when $\visco\not= 0$. It is clear from \fourth\ that
an additional boundary condition is needed to solve the equations.
The additional boundary condition used here is $\delta n(y=0)=0$,
which follows from the analysis of the fields generated by the vortices and
their images near the surface\refs{\brandt}.
Assuming that all perturbations vanish as
$y\rightarrow\infty$,
we obtain
\eqn\densres{\delta n(y)= {\delta H_a\over\phi_0}{\lf^2\over
\lplus^2-\lmin^2}\Big[e^{-y/\lplus}-e^{-y/\lmin}\Big]  .}
where $\lplus(\omega)$ and $\lmin(\omega)$ are two frequency-dependent complex
penetration lengths,
given by,
\eqn\penetrpm{\lambda_{1,2}^2(\omega)={1\over 2}\bigg[\ll_{ac}^2+\delta^2
   \pm\sqrt{(\ll_{ac}^2+\delta^2)^2-{4\ll^2\delta^2\over 1+\ll^2/\lnf^2}
        }\bigg].}
The fields are given by
\eqn\bsemi{\delta B_z(y)={\delta H_a \over
({1\over\lmin^2}-{1\over\lplus^2})}
\bigg[\Big({1\over\lmin^2}-{1\over\lambda^2}-{1\over\lambda_{nf}^2}\Big)
e^{-y/\lplus}-
\Big({1\over\lplus^2}-{1\over\lambda^2}-{1\over\lambda_{nf}^2}\Big)
e^{-y/\lmin}\bigg],}
and
\eqn\efiledvis{E_x(y)=
{-i\omega\over c}{\delta H_a \over
({1\over\lmin^2}-{1\over\lplus^2})}
\bigg[({1\over\lmin^2}-{1\over\lambda^2}+{1\over\lambda_{nf}^2})
\lambda_1e^{-y/\lplus}-
({1\over\lplus^2}-{1\over\lambda^2}+{1\over\lambda_{nf}^2})
\lambda_2e^{-y/\lmin}\bigg].}
In a viscous flux liquid the penetration of the ac field
is governed by two length scales, $\lplus(\omega)$ and
$\lmin(\omega)$.
When $\visco=0$, $\lmin(\omega)=0$ and $\lplus(\omega)=\ll_{ac}(\omega)$.
In this case Eq. \bsemi\ simply reduces to \bfzero .
The role of the new penetration length, $\lmin$,
can be understood by examining  the change $\delta n(y)$
in the vortex density arising from the ac field, given in \densres\ .
The density $\delta n(y)$ reaches its maximum
near $y_0\approx [|\lplus ||\lmin |/(|\lplus |-|\lmin |)]
\ln(|\lplus |/|\lmin |)$.
Since $|\lplus|>>|\lmin|$ for all frequencies of interest here,
$y_0\approx|\lmin|$ and
$\delta n(y_0)\approx (\delta H_a/\phi_0)\lf^2/\lplus^2$.
The density grows from zero at $y=0$ to its maximum value at
$y_0\sim |\lambda_2|$ and then decays to zero over a length $|\lplus| >>
|\lambda_2|$.
In other words both the surface currents associated with
the Meissner response ($\lambda\not=0$) and the
spatially inhomogeneities in the electric field
arising from intervortex interaction ($\visco\not=0$)
impede the build up of an appreciable vortex density
in a surface layer
of width $|\lmin|$. When either $\visco=0$ or $\ll=0$
the width of this surface layer vanishes.
A model that neglects variations on the length scale of order
$\lambda$ (this corresponds to letting $\lambda=0$ in Eqs. (3.13-16))
assumes that $\phi_0\delta n(y)=\delta B_z(y)$
and neglects the surface currents responsible for the jump in the tangential
component of the average magnetic field at the surface.
Similarly, when $\delta=0$ one
neglects spatial inhomogeneities
in the electric field from flux motion at the surface
and the flux-line density again has an unphysical
finite value at the surface, given by
$\delta n(0)=(\delta H_a/\phi_0)\lf^2/(\lf^2+\ll^2)$.
It is also possible to satisify the boundary condition $\delta
n(0)=0$ in a microscopic model that incorporates the interactions of
the vortices among themselves and with their images. This boundary
condition cannot, however, be satisified in the conventional flux flow
model. The hydrodynamic model described here provides a
phenomenlogical description of flux dynamics that can
satisfy this boundary condition
because it incorporates the nonlocalities in the velocity and electric
fields due to intervortex interaction.

Due to the nonlocality induced by the viscous screening the
electrodynamics response of the system needs to be described in terms
of nonlocal response functions. The local real space ac resistivity
and ac permeability are defined by
\eqn\nonlr{E_x(y)=\int dy'\rho_{ac}(y-y',\omega)j_x(y').}
and
\eqn\nonlmu{\delta B_z(y)=\int dy'\mu_{ac}(y-y',\omega)H_a(y').}
In an infinite system the above nonlocal relationships in real space
simply yield the usual linear relationships between the Fourier
components of the fields and currents,
\eqn\fouriere{E_x\qw=\rho_{ac}\qw j_x\qw}
\eqn\fourierb{\delta B_z\qw=\mu_{ac}\qw H_a\qw}
Due to this nonlocality the ac resistivity is not simply related to
the surface impedance.
The voltage drop measured in a transport experiment has to be
evaluated for each specific experimental geometry using realistic
boundary condition, as discussed in \refs{\majum}.
The effect of the viscosity and the crossover between liquid-like and
solid-like behavior should be observable in ac multiterminal
experiments of the type described in \refs{\gammel}.
Similarily, the nonlocality of the permeability could be probed by
measuring local magnetization profiles inside the sample.
The net field penetration in a slab of thickness $2W$ is still described by
the macroscopic permeability defined in \suscep\ ,  which corresponds
to the ${\bf q}=0$ component of $\mu\qw$.
Its expression is given in Eq. (A.2).

To simplify the discussion of the two length scales $\lplus(\omega)$
and $\lmin(\omega)$ and of other electrodynamic properties we
will drop the normal
fluid contribution below. As in the case
$\eta_l=0$, the  normal fluid
contribution introduces an extra crossover at very high frequency
($\omega\sim \omega_{nf}\sim 10^{15}$Hz) and can be easily distinguished
from all the other relaxational modes which occur at frequency scale
at least one order of magnitude less than $\omega_{nf}$.

The new length scale $\lambda_2$ is in magnitude at most of the order
of $\lambda$ at all frequencies. When the static viscosity is small,
i.e. $\delta_0/\lambda << 1$, $\lambda_2$ is always negligible
compared to both
$\lambda_1$ and $\lambda$, as shown in Figs. 3a and
4a. In this case the field penetration is governed by the single
length scale $\lambda_1$, with $\lambda_1\sim\lambda_f$ over
the entire frequency range of interest
\nref\mlondon{ The fact that $\lambda_1$ does
not vanish as $\omega\rightarrow\infty$, but rather approaches
$\lambda$, is the result of having neglected the normal fluid
contribution that becomes dominant at very high
frequencies.}\refs{\mlondon}.
The ac permeability is well approximated by the familiar formula
$\mu(\omega)\approx\lambda_1(\omega)/W \tanh$ $(W/\lambda_1(\omega))$
and is shown in Fig. 3b.
Experiments are typically carried out at frequencies $\omega \leq \omega_f$.
As discussed earlier,
the transition from solid-like to liquid like response
takes place at $\omega\sim\omega_l$.
Since $\omega_l << \omega_f$ in a viscous liquid due to the
small value of the ratio $G/c_L$,
the crossover at $\omega\sim\omega_l$
occurs well into the flux flow regime,
in the sense that $|\lambda_f(\omega)|>>\delta_0$ and conventional
flux flow dominates the response, as shown in Fig. 3b.

The expression used above
for the shear relaxation time $\tau_l$  is simply a
phenomenlogical estimate. In particular the moduli $G$ and $c_L(0)$
appropriate for an entangled polymer glass are not known. An
alternative approach would be to simply consider $\tau_l$ or
$\omega_l$ as a parameter. The effect of viscosity on the ac
permeability will then be observable if the $G/c_L$ is
not too small, while the viscosity is sufficient large. To illustrate this
we show in Fig. 4b-6b the ac
permeability for $\omega_l/\omega_f=0.01$
and $\delta_0/\lambda=1,10,100$. When the viscosity is small
($\delta_0/\lambda=1$), $\lambda_2$ is still negligible at all
frequencies and the response is given by conventional
flux flow with the maximum at $\mu^{''}$ corresponding to $|\lambda_f|\sim
W$ (Fig. 4b). When
$\delta_0/\lambda >> 1 $, $\lambda_2$ can become of order of $\lambda$
and the width of the surface layer where flux penetration is impeded is no
longer negligible. In this regime and for $\omega < \omega_l$ we
can approximate $\lambda_1^2\sim\lambda_{ac}^2+\delta_0^2$ and
$\lambda_2\sim 2\lambda^2\delta_0^2/(\lambda_{ac}^2+\delta_0^2)$.
There is a new crossover at the characteristic frequency
$\omega_\eta$ where $|\lambda_f|\sim\delta_0$ , as shown in Figs. 5a
and 6a. For $\omega << \omega_\eta$ we find $\lambda_1\sim\lambda_f$ while
$\lambda_2\sim\delta_0\sqrt{i\omega/\omega_f}$ is negligible.
For $\omega >> \omega_\eta$ $\lambda_1\sim\delta_0$ and
$\lambda_2\sim\lambda$.
Correspondingly, a second peak develops in $\mu^{''}$ at
$\omega\sim\omega_\eta$ for intermediate viscosity (Fig. 5b).
At very
large viscosity, the crossover in the ac response occurs at
$\omega_\eta$.
To understand this we recall that $|\lambda_2|$ is the
distance from the surface over which the vortex density builds up to
its maximum value, while $|\lambda_1|$ characterizes the decay of the
density from its maximum value to zero within the sample.
If $\omega_\eta << \omega_l $ ( as
it is the case for the parameters of Figs. 6 ),
$|\lambda_1|>>2W$ at all frequencies
($\lambda_1\approx\lambda_f\sim 1/\sqrt{\omega}$ for $\omega <
\omega_\eta$ and $\lambda_1\sim\delta_0$ for
$\omega_\eta < \omega < \omega_l$)
and the ac perturbation simply
cannot penetrate into the sample, unless $|\lambda_2|$ is essentially
zero, as it is the case for
$\omega << \omega_\eta$. The real part of the ac permeability
drops therefore from the value corresponding to complete
penetration ($\mu^{'}=1$) to Meissner response ($\mu^{'}=0.1$) at
$\omega=\omega_\eta$, where $\lambda_2$ becomes comparable to $\lambda$
and the viscous screening discussed above becomes appreciable.

Finally, the surface impedance is obtained by inserting
\efiledvis\ in \surfimp\ ,with the result
\eqn\surfimpa{Z_s(\omega)={-4\pi i\omega\over c^2}
{(\lambda_{ac}^2+\lambda_1\lambda_2)\over (\lambda_1+\lambda_2)}.}
The surface impedance for a superconducting slab of finite
thickness $2W$ in the $y$ direction can be calculated in a
similar way. The result is given in Appendix A.
Again, when $|\lmin|<<\ll<<|\lplus|$ the field penetration is governed by
the longest length scale and Eq. (A.3) can be approximated by
\eqn\surfone{\eqalign{Z_s(\omega)&\approx {4\pi i\omega\over c^2}
\Big[\lplus\tanh(W/\lplus)+{\lmin^3\over\ll^2}
  \Big]\cr
 &\approx {4\pi i\omega\over c^2}\lplus\tanh(W/\lplus)
  .}}
\newsec{ Summary }
We have studied the linear response of a viscous flux-line liquid to
ac perturbations by using a hydrodynamic theory. The flux array is
described as a viscoelastic medium that responds elastically to
perturbations varying on time scales shorter than the characteristic
time $\omega_l^{-1}$ for relaxation of shear stresses, while it
behaves as a viscous fluid on time scales large compare to
$\omega_l^{-1}$. For realistic values of parameters for YBCO, the
characteristic frequency $\omega_l$ is, however, small compared to
the flux flow frequency $\omega_f$. As a result, the crossover from
liquid-like to solid-like behavior occurs well into the flux flow
regime and it generally cannot be detected in the ac permeability
which probes the spatially averaged response of the system. This is
because an ac permeability measurement probes the response of the
flux array to a compression. Both flux liquid and flux solid have
nonvanishing compressional moduli and the values of these moduli in
the two regimes are very similar, in virtue of the very small shear
modulus of the vortex lattice. A signature of the viscous flux-line
liquid that distinguishes it from a flux lattice is the intrinsic
nonlocality of the response. This is responsible for the appearance of
the second penetration length, $\lambda_2$.
It should be possible to probe this
nonlocality by flux imaging experiments, provided the spatial
resolution is sufficiently high to detect variations of length scales
of order of $|\lambda_2|\sim\lambda$.

\bigskip
This work was supported by the National Science Foundation through
grants No. DMR-9112330 and DMR-9217284.

\appendix{A}{AC response for slab geometry}

In this Appendix we describe the response of a superconducting
slab occupying the region $|y|\leq W$ to an ac field
$\zhat\delta H_a e^{i\omega t}$ applied at the sample surfaces.
The local magnetic induction in the superconductor is
given by,
\eqn\bslab{\delta B_z(y)={\delta H_a \over
({1\over\lmin^2}-{1\over\lplus^2})}
\bigg[({1\over\lmin^2}-{1\over\lambda^2}-{1\over\lambda_{nf}^2})
{\cosh(y/\lplus)\over \cosh(W/\lplus)}-
({1\over\lplus^2}-{1\over\lambda^2}-{1\over\lambda_{nf}^2})
{\cosh(y/\lmin)\over \cosh(W/\lmin)}\bigg].}
The ac permeability defined in \suscep\ is
\eqn\susslab{\mu(\omega)=\Big({1\over W}\Big)
{1\over\Big({1\over\lmin^2}-{1\over\lplus^2}\Big)}
\bigg[\Big({1\over\lmin^2}-{1\over\lambda^2}-{1\over\lambda_{nf}^2}\Big)
\lambda_1\tanh(W/\lplus)-
\Big({1\over\lplus^2}-{1\over\lambda^2}-{1\over\lambda_{nf}^2}\Big)
\lambda_2\tanh(W/\lmin)\bigg].}
The surface impedance at one of the boundaries is given by
\eqn\surfimps{\eqalign{Z_s(\omega,y=\pm W)={\mp 4\pi i\omega\over c^2}
{1\over ({1\over\lmin^2}-{1\over\lplus^2})}\bigg\{
&\bigg[({1\over\lmin^2}-{1\over\lambda^2}-{1\over\lambda_{nf}^2})
\lplus \tanh(W/\lplus)\bigg]- \cr
&\bigg[ ({1\over\lplus^2}-{1\over\lambda^2}-{1\over\lambda_{nf}^2})
\lmin \tanh(W/\lmin)\bigg]\bigg\}.}}

\vfill\eject
\listrefs
\listfigs
\end